\documentclass[12pt,a4paper]{article}
\usepackage{fullpage,epsfig,graphics,amsbsy,amssymb,cancel,slashed,mathrsfs}
\usepackage{dsfont}
\usepackage{float}
\usepackage{amsfonts}
\usepackage{psfrag,hyperref}
\usepackage{graphicx,color}
\usepackage{wrapfig}

\usepackage{tikz-cd,tikz}
\usetikzlibrary{calc}
\usetikzlibrary{decorations.pathreplacing}

\newcommand{\be}{\begin{eqnarray}}
\newcommand{\ee}{\end{eqnarray}}
\usepackage{amsmath}
\usepackage{slashed}
\numberwithin{equation}{section}
\usepackage{caption}
\usepackage{subcaption} 
\usepackage[nosort]{cite}

\graphicspath{{./figures/}}

\newcommand{\bea}{\begin{eqnarray}}
\newcommand{\eea}{\end{eqnarray}}  
\newcommand{\nn}{\nonumber}
\newcommand{\Tr}{\textrm{Tr}}
\newcommand{\tr}{\textrm{tr}}

\newcommand{\NN}{\mathcal{N}}

\newcommand{\OO}{{\mathcal O}}

\newcommand{\VV}{\mathcal{V}}

\newcommand{\cp}{\mathbb{C}\textrm{P}^3}

\def\al{\alpha}
\def\om{\omega}

\newcommand{\bsigma}{\Bar{\sigma}}
\newcommand{\bP}{\mathbf{P}}
\newcommand{\bQ}{\mathbf{Q}}
\newcommand{\ii}{i}
\newcommand{\ds}[1]{\mathds{#1}}
\newcommand{\coup}{h} %Coupling Constant
\newcommand{\partitionY}{w}

\usepackage{bbold}

\newcommand{\floor}[1]{\lfloor #1 \rfloor}

\graphicspath{{./figures/}}

\newcommand{\tx}{x}
\newcommand{\ty}{y}

\newcommand{\tB}{B}
\newcommand{\tA}{A}

\newcommand{\tbx}{\mathbf{\tx}}
\newcommand{\tby}{\mathbf{\ty}}

\newcommand{\hatlambda}{\hat{\lambda}}

\begin{document}

%\maketitle
\thispagestyle{empty}
\begin{flushright} \small
UUITP-16/23\\
MIT-CTP/5574
 \end{flushright}
\smallskip
\begin{center} \LARGE
{\bf The ABJM Hagedorn Temperature from Integrability}
 \\[12mm] \normalsize
{\bf  Simon Ekhammar${}^{a,b}$,  Joseph A. Minahan${}^{a,c}$, and Charles Thull${}^{a}$ } \\[8mm]
 {\small\it
 % $^a$
 ${}^a$Department of Physics and Astronomy,
     Uppsala University,\\
     Box 516,
     SE-751\,20 Uppsala,
     Sweden
     
     \smallskip
     \centerline{\it and}
     
     \smallskip
    ${}^b$ Mathematics Department, King’s College London,\\
    The Strand, London WC2R 2LS, UK
     
         \smallskip
     %$^b$
     \smallskip
     \centerline{\it and}
     
     \smallskip
     ${}^c$Center for Theoretical Physics,
     Massachusetts Institute of Technology\\
     Cambridge, MA 02139, USA
     
  }

  \medskip 
   \texttt{ \href{mailto:simon.ekhammar@physics.uu.se,joseph.minahan@physics.uu.se,charles.thull@physics.uu.se}{\{simon.ekhammar, joseph.minahan, charles.thull\}@physics.uu.se}}

\end{center}
\vspace{7mm}
\begin{abstract}
We use the quantum spectral curve to compute  the Hagedorn temperature for ABJM theory in terms of the interpolating function $h(\lambda)$.   At weak coupling we compute this temperature up to eight-loop order, showing that it matches the known tree-level and two-loop results.  At strong coupling we compute the dependence numerically, showing that it is consistent with expectations from supergravity and the plane-wave limit for the four leading terms in the strong coupling expansion, up to an overall shift of the zero-point energy for type IIA string theory on AdS$_4\times \cp$. We conjecture an analytic form for this shift to leading order that is consistent with our numerical results.
\end{abstract}

\eject
\normalsize

\tableofcontents

\section{Introduction}

This is a companion paper to \cite{Ekhammar:2023glu} where we conjectured the  form of the sub-leading contributions to the Hagedorn temperature $T_H$ for planar $\NN=4$ super Yang-Mills at large coupling $\lambda$.  There we improved upon the numerical accuracy in \cite{Harmark:2021qma} of $T_H$ as a function of $g\equiv\frac{\sqrt{\lambda}}{4\pi}$, and demonstrated the consistency of our conjecture.

In this paper we generalize our analysis to the ABJM case \cite{Aharony:2008ug} and make a conjecture for the asymptotic dependence of the Hagedorn temperature. We do so by twisting the AdS$_4$/CFT$_3$ quantum spectral curve (QSC), thus extending the integrability formalism of \cite{Harmark:2017yrv,Harmark:2018red,Harmark:2021qma} from $\mathcal{N}=4$ to ABJM. We then carry out the numerical computation of the Hagedorn temperature in terms of the interpolating function $h(\lambda)$, where $\lambda=N/k$ is the 't Hooft coupling.  We show that these results are consistent with the conjecture. 

The conjecture relies on the supergravity analysis of a light winding mode wrapped around the  time direction in Euclidean AdS$_4$.  The first order correction to the Hagedorn temperature follows from the analysis in \cite{Urbach:2022xzw}.  In this paper we compute the next two orders in the expansion.  The first such correction also depends on the shift of the zero-point energy of the string on the curved AdS$_4\times \cp$ background. We take inspiration from the plane-wave limit to motivate the temperature dependence of this shift. The numerical results indicate that the shift on the full space should be 12/5 the shift for the plane-wave. Interestingly, this shift is 3/4 of what is required for $\NN=4$ super Yang-Mills.  This motivates us to conjecture a shift for general $AdS_{d+1}$ where the shift is proportional to $d$, the dimension of the dual CFT.

We also study the Hagedorn temperature at weak coupling.  Here we are able to compute it up to the eight-loop order, which goes well beyond the previous limits.  Our result at tree-level agrees with the  previously computed value in \cite{Nishioka:2008gz}.  Our result at two loops agrees with the value of \cite{Papathanasiou:2009en}, once a subtle factor of two is taken into account.

The rest of this paper is structured as follows.  In section \ref{subsec:WeakCoupling} we discuss the Hagedorn temperature at tree-level and at the two-loop level.  In section \ref{subsec:Sugra} we compute the supergravity corrections to the Hagedorn temperature by mapping the problem of a light winding mode on the compactified time direction in AdS$_{d+1}$ to a perturbed radially symmetric harmonic oscillator in $d$ dimensions.  In section \ref{subsec:ppwave} we discuss the plane-wave limit of the type IIA string theory and show that it gives a first order correction to the Hagedorn temperature which matches the supergravity result.  We also show that the shift to the zero-point energy effects the next correction, but not the following one.  In section \ref{sec:ABJMqsc}, we outline the general form of the twisted AdS$_4$ QSC before specializing for the ABJM Hagedorn temperature. In section \ref{sec:PertQSCSol} we give the perturbative solution of the QSC to order $h^8$ and in \ref{sec:StrongCoupling} we discuss the numerical solution for strong coupling.

{\it Note added:} As this paper was in the final stages of preparation \cite{Bigazzi:2023hxt} appeared which also derives \eqref{THd} and   \eqref{ppABJMbeta} in this paper.  Furthermore, their paper gives a world-sheet argument for the conjectured value of the zero-point shift given in \eqref{deltcconj}.

\section{The Hagedorn temperature at weak and strong coupling}

\subsection{Weak coupling}\label{subsec:WeakCoupling}
At zero coupling the partition function of ABJM can be computed by stringing beads together into a single-trace partition function and then exponentiating to multi-trace contributions. In ABJM we have two different types of ``half-beads" coming from the two basic sets of fields, $(\phi_a,\psi^a)$ and $(\overline{\phi}{}^a,\overline{\psi}_a)$, which sit in the $(N,\overline{N})$ and $(\overline{N},N)$ of the gauge groups, respectively. Here $\phi$'s are bosons, $\psi$'s fermions and the index $a$ labels the (anti-)fundamental representation of $\mathfrak{su}_4$. Due to the structure of the gauge groups this means that the chain for a single-trace operator has to be alternating.

The tree-level Hagedorn temperature was first computed in \cite{Nishioka:2008gz}. We will here briefly recall how this computation is carried out. Let us define
\begin{equation}
    Z = \tr\, e^{-\beta H_0}  = \tr \, \partitionY^{2 D_0}\,,
\end{equation}
 where $\partitionY\equiv e^{-\frac{1}{2T}}$ and $D_0$ is the dilatation operator for the free theory. Then the single-bead partition function is 
\be
    Z_{bead}(\partitionY)=(Z_\phi(\partitionY)+Z_\psi(\partitionY))^2\,,
\ee
where
\be
    Z_\phi=\frac{4\partitionY(1+\partitionY^2)}{(1-\partitionY^2)^2}\,\quad{\rm and}\quad Z_\psi=\frac{8\partitionY^2}{(1-\partitionY^2)^2}\,.
\ee
Hence, it follows that
\be
     Z_{bead}=\left(\frac{4\partitionY}{(1-\partitionY)^2}\right)^2\,.
\ee
The tree level value of the Hagedorn temperature is then found by setting $Z_{bead}(\partitionY_H)=1$, from which it follows that \begin{equation}\label{wH}
\partitionY_H=3-2\sqrt{2}\,,    
\end{equation}
and so
\begin{align}\label{eq:Thtree}
    T_H&=\frac{1}{2\log(3+2\sqrt{2})}\approx 0.283648\,.
\end{align}

The leading correction to the Hagedorn temperature at weak coupling in ABJM theory was computed in \cite{Papathanasiou:2009en} from the two-loop dilatation operator of \cite{Zwiebel:2009vb,Minahan:2009te}. They found that\footnote{We have included a subtle factor 2 compared to the result stated in \cite{Papathanasiou:2009en}. This factor arises because  the ``pendant" should include a contribution from both $\langle D_2(x)\rangle$ and $\langle \overline{D}_2(x)\rangle$, where $\langle D_2(x)\rangle=\Tr_{\VV^1\otimes\overline{\VV}^1\otimes\VV^1}x^\Delta D_2$ and
$\langle \overline{D}_2(x)\rangle=\Tr_{\overline{\VV}^1\otimes\VV^1\otimes\overline{\VV}^1}x^\Delta D_2$.  See \cite{Papathanasiou:2009en} for details.  We thank G. Papathanasiou and M. Spradlin for correspondence on this isssue.}
\begin{equation}\label{eq:Th1}
    \frac{\delta T_H}{T_H}=\frac{\lambda^2}{\sqrt{2}}\langle D_2(\partitionY_H^{-2})\rangle=4\lambda^2(\sqrt{2}-1).
\end{equation} 

In section \ref{sec:PertQSCSol} we will verify the tree-level and two-loop results, and also find the next three terms in the perturbative expansion using the QSC.

\subsection{Strong coupling and supergravity}\label{subsec:Sugra}

In this subsection we generalize the discussion in \cite{Ekhammar:2023glu} to theories whose supergravity dual is $AdS_{d+1}$.  In this case the Euclidean $AdS$ metric is given by
\be
ds^2=(1+R^2)d\tau^2+\frac{dR^2}{1+R^2}+R^2 d\Omega_{d-1}^2\,,
\ee
where we have set the radius  to $1$ and 
made the identification $\tau\equiv\tau+\beta$.  We then assume that a string winds once around the $\tau$ direction such that the world-sheet fermions have anti-periodic boundary conditions.  These boundary conditions shift the zero-point energy of the string to a nonzero value, $C=C_0+\Delta C$ where $C_0=-2/\al'$ is the shift in flat space.  If $\beta$ is tuned properly this winding mode is very light and we can use the supergravity approximation.  The winding mode is a scalar field $\chi$ and its contribution to the action is
\be 
\int d^{d+1}X \sqrt{g}\left(\nabla^\mu\chi\nabla_\mu\chi+m^2(R)\chi^2\right)\,,
\ee
where $m^2(R)$ is the radial dependent mass-function
\be
m^2(R)=(1+R^2)\left(\frac{\beta}{2\pi\al'}\right)^2+C\,.
\ee
Choosing $\chi$ to be massless means finding a $\tau$ independent normalizable solution to the equations of motion.  Assuming that $\chi$ only has $R$ dependence then leads to the equation
\be\label{EOM}
&&-\frac{1}{2}\frac{1}{R^{d-1}}\frac{d}{dR} R^{d-1}\frac{d}{dR} \chi(R)+\frac{1}{2}\left(\frac{\beta}{2\pi\al'}\right)^2R^2\,\chi(R)+\Delta H\,\chi(R)\nn\\
&&\qquad\qquad\qquad\qquad\qquad\qquad\qquad\qquad=-\frac{1}{2}\left(C+\left(\frac{\beta}{2\pi\al'}\right)^2\right)\chi(R)\,,
\ee
where 
\be\label{Hpert}
\Delta H =-\frac{1}{2}\frac{1}{R^{d-1}}\frac{d}{dR} R^{d+1}\frac{d}{dR}\,.
\ee
We want a solution to \eqref{EOM} that is normalizable and hence falls off as $R\to\infty$.  Hence, solving \eqref{EOM}  is equivalent to solving the radial eigenvalue equation for  a perturbed $d$-dimensional harmonic oscillator with frequency $\om=\frac{\beta}{2\pi\al'}$ and energy 
\be
E=-\frac{1}{2}\left(C+\left(\frac{\beta}{2\pi\al'}\right)^2\right)\,.
\ee

To lowest order we drop $\Delta H$ and set $C=-2/\al'$.  The ground-state energy for the harmonic oscillator then satisfies
\be\label{Eeq0}
E=\frac{1}{2}\left(\frac{2}{\al'}-\left(\frac{\beta}{2\pi\al'}\right)^2\right)=\frac{d}{2}\,\om=\frac{d}{2}\,\frac{\beta}{2\pi\al'}\,,
\ee
which has the solution
\be
\beta=\pi\al'\left(\sqrt{\frac{8}{\al'}+d^2}-d\right)\,.
\ee
Hence, to leading order we have that
\be
T_H=\frac{1}{\beta}=\frac{1}{2\sqrt{2\al'}}+\frac{d}{8\pi}+\dots\,.
\ee
The first order correction matches that in \cite{Urbach:2022xzw}.

For the next two orders we  use first and second order perturbation theory for the perturbed oscillator, as well as the correction $\Delta C$.  As described in \cite{Ekhammar:2023glu} and in the next subsection, an analysis of the plane-wave Hagedorn temperature suggests that 
\begin{equation}\label{zptsh}
    C=-\frac{2}{\alpha'}+\frac{\beta^2}{2\pi^2\alpha'}\Delta c+\mathcal{O}(\alpha').
\end{equation}
 Furthermore, the correction to the energy is 
\begin{align}
    \Delta E&=\langle\psi_0|\Delta H|\psi_0\rangle+\frac{\langle\psi_0|\Delta H|\psi_2\rangle\langle\psi_2|\Delta H|\psi_0\rangle}{-2\omega}+\frac{\langle\psi_0|\Delta H|\psi_4\rangle\langle\psi_4|\Delta H|\psi_0\rangle}{-4\omega}\\
    &=\frac{d(d+2)}{8}+0-\frac{d(d+2)}{32 \omega}\,,
\end{align} 
where the relevant normalized wave-functions are
\begin{align*}
    \psi_0(R)&=\sqrt{\frac{2}{\Gamma(d/2)}}\omega^{d/4}e^{-\frac{1}{2}\omega R^2}\,,\\
    \psi_2(R)&=\sqrt{\frac{2}{\Gamma(d/2+1)}}\omega^{d/4+1}\left(2R^2-\frac{d}{\omega}\right)e^{-\frac{1}{2}\omega R^2}\,,\\
    \psi_4(R)&=\frac{1}{\sqrt{\Gamma(d/2+2)}}\omega^{d/4+2}\left(R^4-\frac{d+2}{\omega}R^2+\frac{d(d+2)}{4\omega^2}\right)e^{-\frac{1}{2}\omega R^2}\,.
\end{align*}
Hence the equation in \eqref{Eeq0} becomes
\begin{equation}
    \frac{1}{2}\left(\frac{2}{\alpha'}-\left(\frac{\beta}{2\pi\alpha'}\right)^2-\frac{\beta^2}{2\pi^2\alpha'}\Delta c\right)=\frac{d}{2}\frac{\beta}{2\pi\alpha'}+\frac{d(d+2)}{8}-\frac{d(d+2)}{32 }\frac{2\pi\alpha'}{\beta}\,,
\end{equation} 
which we can rearrange into the form
\be\label{betaeq2}
\frac{\beta^2}{4\pi\al'}=2\pi-\frac{d}{2}\beta-\frac{\beta^2\Delta c}{2\pi}-\frac{d(d+2)}{4}\pi\al'+\frac{d(d+2)\pi^2(\al')^2}{8\,\beta}\,.
\ee
From this we find the Hagedorn temperature 
\begin{equation}\label{THd}
    T_H(\alpha')=\frac{1}{2\pi\sqrt{2\alpha'}}+\frac{d}{8\pi}+\frac{d(d+1)+8\Delta c}{16\sqrt{2}\pi}\sqrt{\alpha'}+\frac{d(d+2)(4d-1)}{256\pi}\alpha'+\mathcal{O}\left((\alpha')^{3/2)}\right).
\end{equation}
Based on the numerical results in \cite{Ekhammar:2023glu} and in section \ref{results} we conjecture the value of $\Delta c$ in \eqref{deltcconj}.

Applying \eqref{THd} to ABJM, we set $d=3$ and use the dictionary $\alpha'=\frac{1}{\pi\sqrt{2\hat\lambda}}$, where $\hat\lambda\equiv\lambda-\frac{1}{24}$ is the shifted 't Hooft parameter \cite{Bergman:2009zh} \footnote{At the highest order shown in \eqref{THd} it does not matter if we use $\lambda$ or $\hat\lambda$.  However, later when we discuss numerical results we will go to higher order in $\hat\lambda^{-1}$, where $\hat\lambda$ is the more natural parameter.  We thank N. Bobev for comments about this.}.  Hence, we find 
\be\label{THconj}
    T_H(\hat\lambda)&=&\frac{(\hat\lambda/2)^{1/4}}{2\sqrt{\pi}}+\frac{3}{8\pi}+\frac{3+2\Delta c}{8\pi^{3/2}}\left(\frac{\hat\lambda}{2}\right)^{-1/4}+\frac{165}{512\pi^2}\left(\frac{\hat\lambda}{2}\right)^{-1/2}+\mathcal{O}(\hat\lambda^{-3/4})\nn\\
    &=&0.237212\hat\lambda^{1/4}+0.119366+(0.0800874+0.0533916\Delta c)\hat\lambda^{-1/4}\nn\\
    &&\qquad\qquad\qquad\qquad+0.0461774\hat\lambda^{-1/2}+\mathcal{O}(\hat\lambda^{-3/4})\,.
\ee
Significantly, the structure of the proposed zero-point shift in \eqref{zptsh} means that the $\lambda^{-1/2}$ coefficient is independent of the shift.

\subsection{Strong coupling and the plane-wave limit}\label{subsec:ppwave}
In \cite{Ekhammar:2023glu} we used the plane-wave limit of type IIB string theory to gain some insight on the structure of the zero-point energy corrections.  We also argued that the first correction to the Hagedorn temperature can be extracted from the growth of states in the plane-wave limit.  In this subsection we will find a similar structure when taking the plane-wave limit of type IIA string theory on AdS$_4\times \cp$.

The type IIA plane-wave metric and flux are given by \cite{Sugiyama:2002tf,Hyun:2002wu}
\be
&&ds^2=-2dx^+dx^-A(x^I)dx^+dx^++\sum_{I=1}^8 dx^Idx^I\,,\nn\\
&&\qquad\qquad F_{+123}=\mu\,,\qquad F_{+4}=\frac{\mu}{3}\,,
\ee
where the prefactor $A(x^I)$ is given by
\be
A(x^I)=\left(\frac{\mu}{3}\right)^2\sum_{I=1}^4x^Ix^I+\left(\frac{\mu}{6}\right)^2\sum_{I=5}^8x^Ix^I\,.
\ee
For the plane-wave reduction from AdS$_4\times \cp$ one has that $\mu=3/\sqrt{2}$.

The free energy for the non-interacting type IIA plane-wave string theory was computed in \cite{Hyun:2003ks}. By looking for a divergence they found that the Hagedorn temperature is found by solving the equation
\begin{multline}\label{ppbeta}
    \frac{\beta^2}{4\pi\alpha'}=2\pi-\frac{\beta\mu}{\sqrt{2}}+\frac{5\beta^2\mu^2}{36\pi}\log(2)\\
    +\sum_{k=2}^\infty \frac{(-1)^k\Gamma(k-\frac{1}{2})}{\Gamma(-\frac{1}{2})\Gamma(k+1)}16\pi(2^{2k-2}-1)(1+2^{-2k})\zeta(2k-1)\left(\frac{\beta\mu}{6\sqrt{2}\pi}\right)^{2k}\,.
\end{multline} 
Note that the shift in the string zero-point energy has the form in \eqref{zptsh}.

If we use the flux for the AdS$_4\times \cp$ reduction \cite{Nishioka:2008gz}, we find
\be\label{ppABJMbeta}
 \frac{\beta^2}{4\pi\alpha'}=2\pi-\frac{3}{2}\,\beta+\frac{5\beta^2}{8\pi}\log(2)+\dots\,.
\ee
The term linear in $\beta$ matches the corresponding term in \eqref{betaeq2}, hence the growth of states in the plane-wave gives the same first-order correction to the flat space Hagedorn temperature as computed from supergravity.  The same result can be found by generalizing the discussion in \cite{Bigazzi:2023oqm}.  If we compare the next term in \eqref{ppABJMbeta} with \eqref{betaeq2} then we see that the plane wave gives $\Delta c=-\frac{5}{4}\log(2)$.

\section{The ABJM QSC}\label{sec:ABJMqsc}
The ABJM QSC was first formulated in \cite{Cavaglia:2014exa}, building on the AdS$_4$/CFT$_3$ TBA  \cite{Bombardelli:2009xz,Gromov:2009at}. The full structure of the QSC is explained in detail in \cite{Bombardelli:2017vhk}. The main object for us will be the ``spinorial" Q-functions $Q_{a|i},\ Q^{a}{}_{|i},\ a,i=1,2,3,4$, constrained to satisfy $\det Q^{a}{}_{|i}=\det Q_{a|i}=-1$. As opposed to the Q-functions that appear in the study of spin chains, the $Q_{a|i}$ and $Q^{a}{}_{|i}$ are functions with branch points at $\pm 2 \coup -\frac{\ii}{2}-\ii n$, $n\in \mathbb{Z}_{\geq 0}$. Here $\coup(\lambda)$ is the integrability coupling constant whose conjectured relation to $\lambda$ is given by  \cite{Gromov:2014eha} \begin{equation}\label{eq:hFromTHooft}
   \lambda=\frac{\sinh(2\pi h)}{2\pi}{}_3F_2\left(\frac{1}{2},\frac{1}{2},\frac{1}{2};1,\frac{3}{2};-\sinh^2(2\pi h)\right)\,.
\end{equation}
In particular, for  $\lambda\ll 1$ we have that 
\begin{equation}
h(\lambda)=\lambda -\frac{\pi^2}{3} \lambda^3+\frac{5\pi^4}{12}\lambda^5-\frac{893\pi^6}{1260}\lambda^7+\OO(\lambda^9)
\end{equation} 
while for $\lambda\gg 1$ the behavior is 
\be\label{largelam}
    h(\lambda)&=&\sqrt{\frac{1}{2}\left(\lambda-\frac{1}{24}\right)}-\frac{\log(2)}{2\pi}+\OO\left(e^{-\pi\sqrt{8\lambda}}\right)\nn\\
    &=&\sqrt{\frac{1}{2}\,\hat\lambda}-\frac{\log(2)}{2\pi}+{\OO}\left(e^{-\pi\sqrt{8\hat\lambda}}\right)\,.
\ee
At strong coupling the shifted 't Hooft parameter $\hat\lambda$ is the more natural parameter \cite{Bergman:2009zh}, as can be seen for example when localizing ABJM \cite{Bobev:2022eus,Bobev:2022jte}.

\begin{figure}
    \centering
    \begin{subfigure}{0.3\textwidth}
        \includegraphics[width=\textwidth]{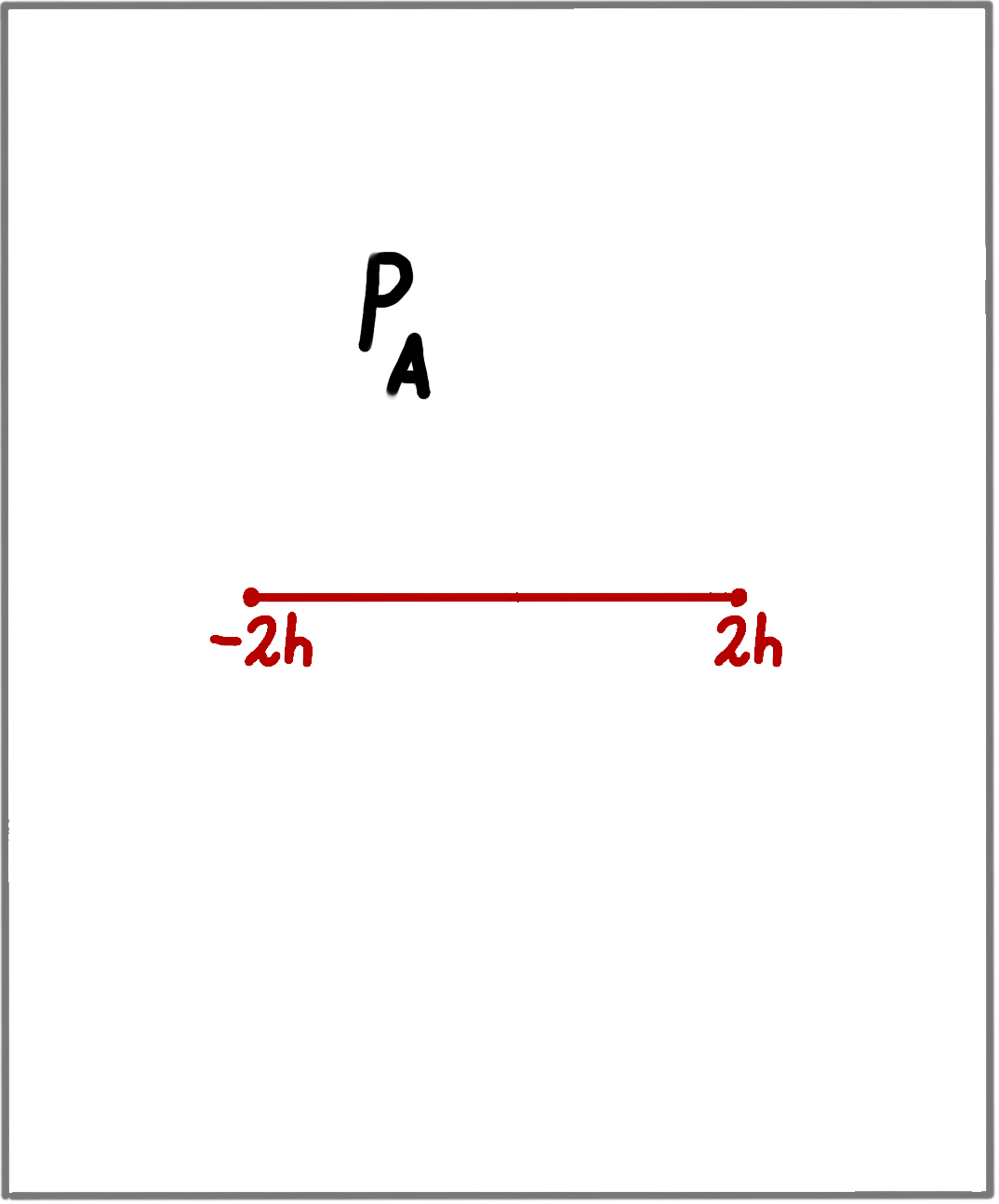}
    \end{subfigure}
    \hfill
    \begin{subfigure}{0.3\textwidth}
        \includegraphics[width=\textwidth]{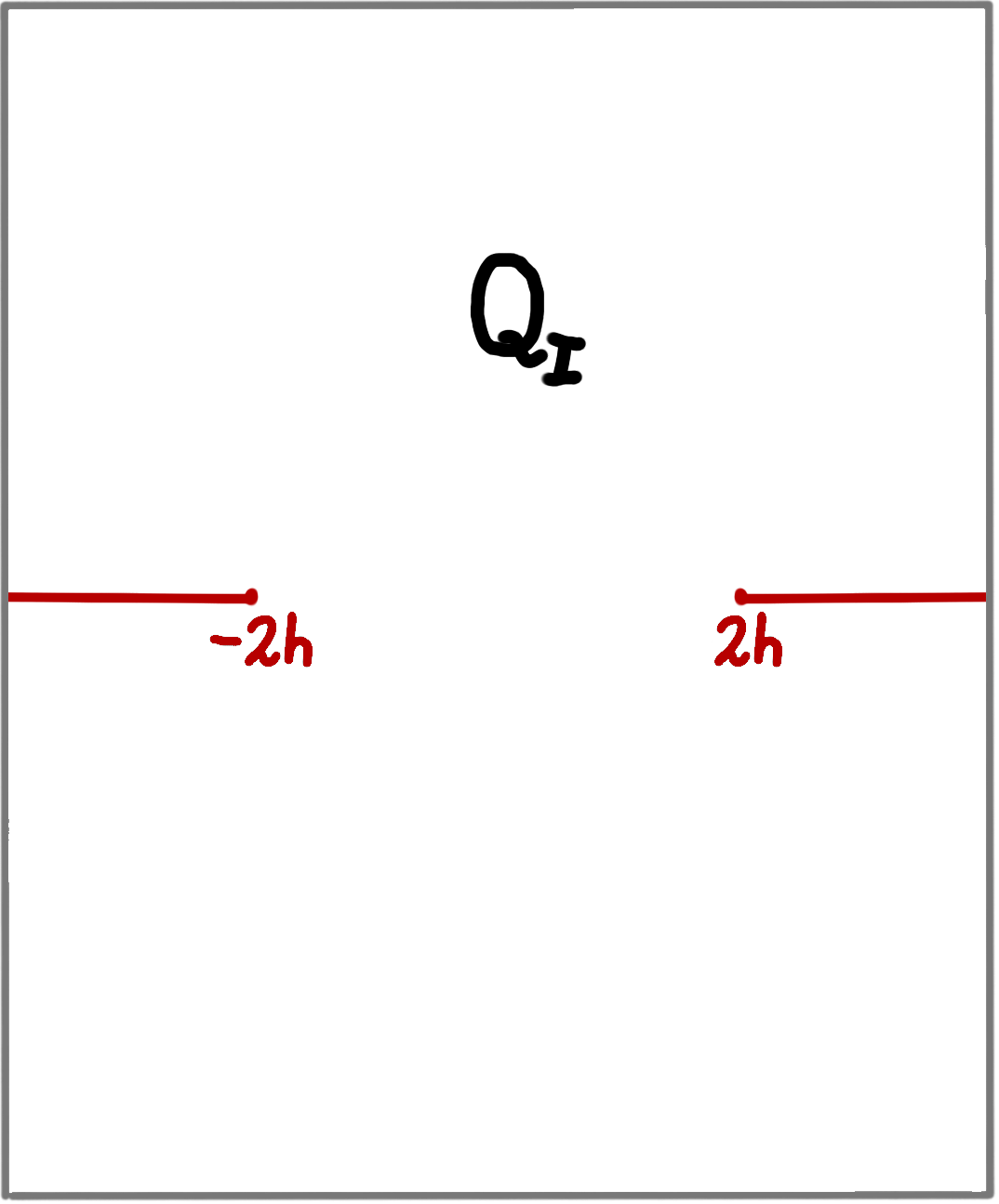}
    \end{subfigure}
    \hfill
    \begin{subfigure}{0.3\textwidth}
        \includegraphics[width=\textwidth]{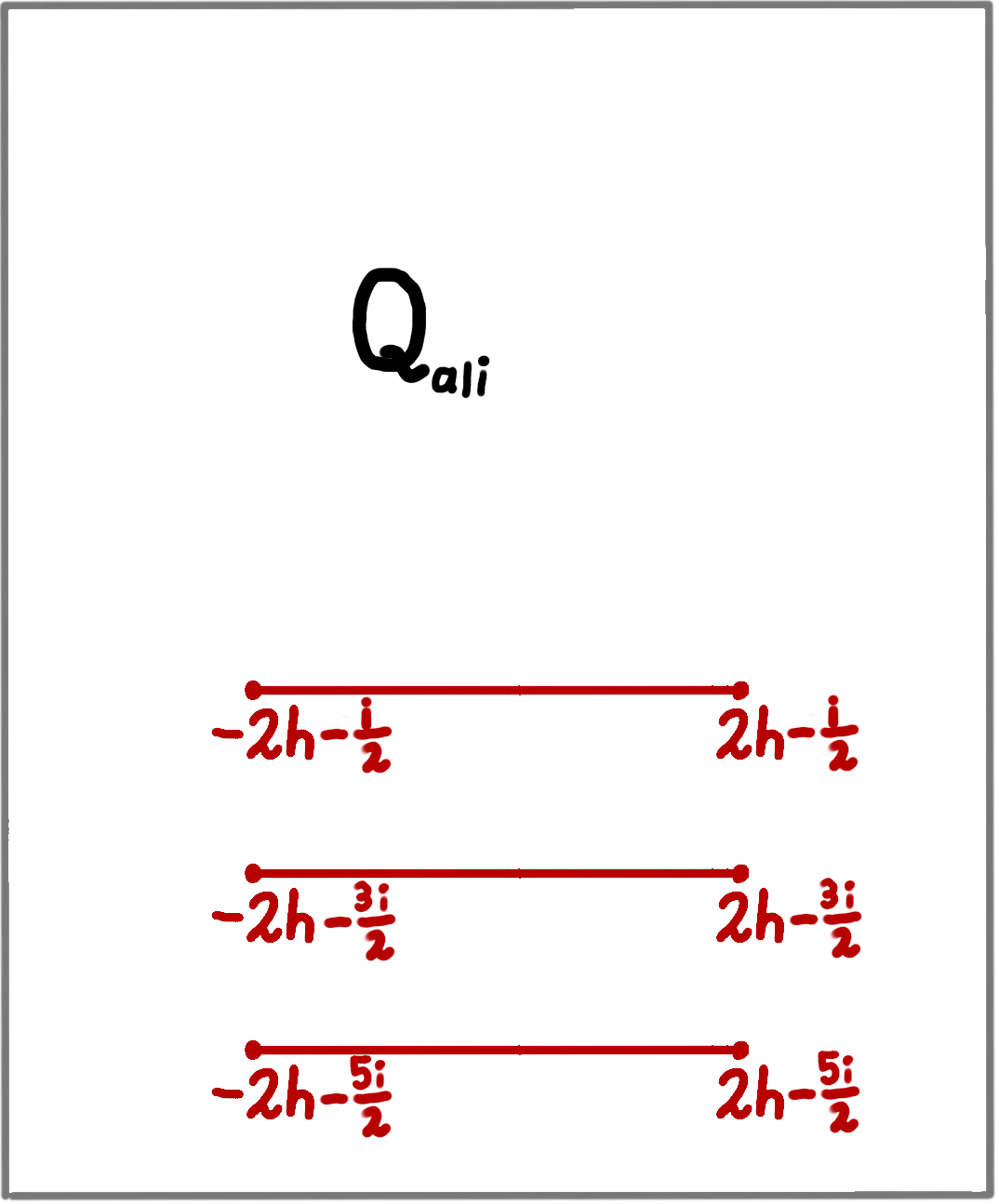}
    \end{subfigure}
    \caption{Cut structure of the $\bP$-, $\bQ$- and $Q_{a|i}$-functions.}\label{fig:Cutstructure}
\end{figure}

$Q_{a|i}$ and $Q^{a}{}_{|i}$ are not independent but are related through
\begin{align}\label{eq:InverseEq}
    &-\kappa^{ij}Q^{a}{}_{|j}Q_{a|k} = \delta^{i}_j\,,
    &
    &\kappa^{ij}Q^{a}{}_{|i}Q_{b|j} = \delta^{a}_{b}\,,
\end{align}
with $\kappa^{ij} = (-1)^{i}\delta^{i+j,5}$. Two more important bilinear expressions are
\begin{align}\label{eq:PQfromQai}
    &\bP_A = -\frac{1}{2}Q^+_{a|i}\kappa^{ij}\bsigma^{ab}_A Q_{b|j}^-\,,\\
    \label{eq:PQfromQai2}
    &\bQ_I = -\frac{1}{2}(Q^{a}{}_{|i})^+ \,\overline{\Sigma}_I^{ij}\, Q_{a|j}^-\,,
\end{align}
where $A=1,\dots,6$ and $I=1\dots,5$.  Explicit expressions for $\Bar{\sigma}^{ab}_A$, $\overline{\Sigma}_I^{ij}$, as well as  other matrices that  appear throughout the main text are collected in Appendix~\ref{app:Conventions}.
The $\bP_A$ and $\bQ_I$ have particularly simple analytic properties. $\bP_A$ must be a function with a short cut while $\bQ_I$ must be a function with a long cut. We show these cuts in figure \ref{fig:Cutstructure}.

To study the Hagedorn temperature we need to twist the QSC. This implies that  the Q-functions will have exponential asymptotics, as opposed to the polynomial asymptotic featured in the spectral problem. We can turn on a total of $3+2=5$ twists for $\mathfrak{osp}(6|4)$. We will use the notation $\mathbf{x}=\{x_1,x_2,x_5\}$ for the $\mathfrak{so}_6$ twists and $\tby=\{y_1,y_2\}$ for the $\mathfrak{sp}_4\simeq \mathfrak{so}_5$ twists. In the fully twisted case the asymptotics of $Q_{a|i}$ for large $u$, up to constants prefactors, are given by

\begin{align}\label{eq:TwistedQaiAsymptotics}
    &Q_{a|i} \sim \mathbf{x}^{\ii u \omega_a} \,\mathbf{y}^{-\ii u\nu_i}\,, 
    &
    &Q^{a}{}_{|i} \sim \tbx^{-\ii u \omega_a} \, \tby^{-\ii u \nu_i}\,,
    &
    &\omega = \begin{pmatrix}
        +++\\
        +--\\
        -+-\\
        --+
    \end{pmatrix}\,,
    \quad
    \nu =\begin{pmatrix}
        ++\\
        +-\\
        -+\\
        --
    \end{pmatrix}\,,
\end{align}
where $\pm = \pm\frac{1}{2}$ so that, for example, $\mathbf{x}^{+++} = x^{\frac{1}{2}}_1 x_2^{\frac{1}{2}}x_5^{\frac{1}{2}}$. Using \eqref{eq:PQfromQai} this translates in the fully twisted case into
\begin{align}\label{eq:FullyTwistedPQ}
    &\bP_{A} \sim \tx_A^{\ii u}\,,
    &
    &\bQ_I \sim \ty^{-\ii u}_I\,,
\end{align}
with the identifications $x_3=\frac{1}{x_2},x_4=\frac{1}{x_1},x_6=\frac{1}{x_5}, y_3=\frac{1}{y_2}$ and $y_4= \frac{1}{y_1}$. These pairings are a consequence of our choice of basis. We emphasize that \eqref{eq:FullyTwistedPQ} is valid only if we assume that all $x_a$ and $y_i$ are independent, otherwise one should include additional powers of $u$.

Throughout this article, we will work in the symmetric sector.  In this case we can identify
\begin{align}\label{eq:SymmetriSectorQcs}
    Q^{a}{}_{|i} = -\kappa^{ab}Q_{b|j}\ds{K}^j_i\,,
\end{align}
which enforces $\tx_5=1$. Furthermore, in this sector we also have
$\bP_{5} = \bP_{6}$ and $\bQ_{5} = 0$.

\subsection{The AdS$_4$ Hagedorn QSC}
We will from now on turn off all R-symmetry twists, thus setting $\tx_a = 1$. We fix the remaining twists to
\begin{equation}\label{eq:yDefinition}
    \ty_1^{-\ii u}=\ty_2^{-\ii u} \equiv \ty^{-\ii u}= e^{-\pi u} e^{-\frac{\ii u}{2\,T_H}}\,.
\end{equation}
The QSC naturally computes the Witten index.  The factor of $e^{\pi u}$ twists the trace in such a way to  give instead the standard thermal partition function. Since there is no twisting of the $R$-symmetry, the asymptotics of \eqref{eq:TwistedQaiAsymptotics} picks up additional powers of $u$, resulting in
\begin{equation}\label{eq:AsymptoticQaiNoRTwist}
    Q_{a|i} \sim \begin{pmatrix}
        \ty^{-\ii u} & u & 1 & \ty^{\ii u}\\
        \ty^{-\ii u}u^2 & u^3 & u^2 & \ty^{\ii u}u^2\\
        \ty^{-\ii u}u & u^2 & u & \ty^{\ii u}u\\
        \ty^{-\ii u}u^3 & u^4 & u^3 & \ty^{\ii u}u^3
     \end{pmatrix}_{a|i}\,,
\end{equation}
and 
\begin{align}\label{eq:AsymptoticQP}
    &\bQ_I \simeq \begin{pmatrix}
        \tB_{1}\,\ty^{-\ii u}\, u\\
        \tB_{2}\,\ty^{-\ii u}\\
        \tB_{3}\,\ty^{\ii u}\, u\\
        \tB_{4}\,\ty^{\ii u}\\\
        0
    \end{pmatrix}_I\,,
    &
    &\bP_{A} \simeq \begin{pmatrix}
        \tA_{1}\, u\\
        \tA_{2} \\
        \tA_{3}\, u^4 \\
        \tA_4\, u^3 \\
        \tA_{5}\, u^2 \\
        \tA_{5}\, u^2
    \end{pmatrix}_{A}\,.
\end{align}
The prefactors $\tA$ and $\tB$ satisfy the relations
\begin{align}\label{eq:tAtBConditions}
    &-\tA_2 \tA_3 = -\frac{1}{4}\tA_1 \tA_4 = \frac{1}{3}\tA_5 \tA_6 = \frac{4}{3} \cosh^4{\frac{1}{4T_H}}\,,
    &
    &\tB_1 \tB_4 = \tB_2\tB_3 = \ii \coth^3{\frac{1}{4T_H}}\,.
\end{align}

Q-functions are not uniquely specified, rather we can transform our entire Q-system by acting on $Q_{a|i}$ with linear transformations acting on either the $a$ or $i$ index. These transformations are usually called $H$-rotations. These transformations are further restricted by \ref{eq:SymmetriSectorQcs} and by
\eqref{eq:AsymptoticQaiNoRTwist}. We found it convenient to fix part of this freedom by imposing that
\begin{align}
    &A_{1} = A_2 = 1\,,
    &
    &A_5=A_6 = 2 \cosh^2{\frac{1}{4T_H}}\,,
    &
    &B_{1} = B_{2} = 8 \cosh^3{\frac{1}{4T}}\,.
\end{align}

The Q-system also exibits parity symmetry, that is, after sending $u\rightarrow -u$ the Q-system is equivalent to the original one up to symmetry transformations. Explicitly
\begin{align}\label{eq:ParityQ}
    &\underline{Q}_{a|i} = \mathbb{g}_{a}{}^{b}\,Q_{b|j}\,(\Omega^{j}{}_{i})^+
    &
    &\underline{\bP}_A = \textrm{diag}(-1,1,1,-1,1,1)_A{}^B\bP_B\,,
\end{align}
where we have used the notation $f(-u) = \underline{f}(u)$. Here $\mathbb{g}$ is a constant matrix while $\Omega^{i}{}_j = (\Omega^{i}{}_{j})^{[4]}$ is in general a more complicated function with branch cuts. However, at large $u$ it is simply a constant. With our specific choice of gauge 
\begin{align}\label{eq:AsymptoticOmega}
    &\mathbb{g} = \begin{pmatrix}
        \ii & 0 & 0 & 0 \\
        0 & \ii & 0 & 0 \\
        0 & 0 & -\ii & 0 \\
        0 & 0 & 0 & -\ii 
    \end{pmatrix}\,,
    &
    &\Omega \simeq \begin{pmatrix}
        0 & 0 & 0 & -\frac{1}{8 \sinh^3{\frac{1}{2 T_H}}}\\
        0 & \ii & 0 & 0\\
        0 & 0 & -\ii & 0\\
        8 \sinh^3{\frac{1}{2T_H}} & 0 & 0 & 0\\
    \end{pmatrix} \,.
\end{align}
The interesting part of these matrices is the distribution of non-zero entries rather than their exact value, which is dependent on normalization.

Since $\bP_A$ is a function with one short cut, we can parameterize it as 
\begin{equation}\label{Pansatz}
\bP_A = \sum_{m=n_A}^\infty \frac{c_{A,m}}{x(u)^m}\,,  
\end{equation}
where $x(u)$ is the Zhukovsky variable
\be
x(u)=\frac{1}{2\coup}\left(u+\sqrt{u+2\coup}\sqrt{u-2\coup}\right)\,,
\ee
and $n_A$ is an $A$ dependent integer.
Using the parity symmetry and the asymptotic behavior described in \eqref{eq:AsymptoticQP} and \eqref{eq:tAtBConditions}, our {\it ansatz} in \eqref{Pansatz} takes the form
\begin{subequations}
\label{eq:AnsatzP}
\begin{align}
    \bP_1 &=x \coup+\sum_{n=1}^\infty c_{1,2n-1}\frac{\coup^{2n-1}}{x^{2n-1}}\,,\\
    \bP_2&=1+\sum_{n=1}^\infty c_{2,2n}\frac{\coup^{2n}}{x^{2n}}\,,\\
    \bP_3 &=-\frac{4\cosh^4\left(\frac{1}{4T_H}\right)}{3}(x \coup)^4\left(1+\sum_{n=2}^\infty c_{3,2n-4}\frac{\coup^{2(n-4)}}{x^{2n}}\right)\,,\\
    \bP_4 &=-\frac{16\cosh^4\left(\frac{1}{4T_H}\right)}{3}(x \coup)^3\left(1+c_{4,-1}\frac{\coup^{-2}}{x^2}+\sum_{n=2}^\infty c_{4,2n-3}\frac{\coup^{2(n-3)}}{x^{2n}}\right)\,,\\
    \bP_5=\bP_6&=2\cosh^2\left(\frac{1}{4T_H}\right)(x \coup)^2\left(1+c_{5,0}\frac{\coup^{-2}}{x^2}+\sum_{n=2}^\infty c_{5,2n-2}\frac{\coup^{2(n-2)}}{x^{2n}}\right)\,,
\end{align}
\end{subequations}
where powers of $\coup$ have been inserted so that $c_{A,n} \simeq \mathcal{O}(\coup^0)$. We have also used the $H$-symmetry to eliminate a possible $x^2$ term in $\bP_4$.

Finally, to close the QSC we supplement it with the so-called gluing conditions. As a function with short cuts $\bQ_I$ is an upper half-plane analytic function with a ladder of cuts $(-2h - \ii n,2h - \ii n),n\in \mathbb{Z}_{\geq0}$ in the lower half-plane. In order to restore the long cut structure described previously, the analytic continuation of $\bQ_{I}$ around the branch-points on the real axis, denoted by $\widetilde{\bQ}_I$, must be a lower half-plane analytic function with a ladder of cuts in the upper half-plane. Luckily, we can easily construct such a function; namely $\bQ_{I}(-u) = \underline{\bQ}_I$. We demand that $\widetilde{\bQ}_{I} = \mathcal{L}_{I}{}^{J}\underline{\bQ}_{J}$ where $\mathcal{L}_I^{J}$ is a matrix with periodicity ${\mathcal{L}_I{}^{J}}^{[2]} = \mathcal{L}_I{}^{J}$. To fix the form of $\mathcal{L}_I{}^{J}$ one should utilize the so-called $\bQ\, \omega$-system as done in \cite{Bombardelli:2017vhk} for the spectral problem. We will not perform this analysis but rather argue the general form of $\mathcal{L}_I{}^J$ from symmetry.

To constrain $\mathcal{L}_I{}^J$ we use the asymptotic properties of $\bQ_I$. At $u=\ii \infty$ we have that $\bQ_{1}>\bQ_2>\bQ_3>\bQ_4$, and demand that the same ordering is true for $\widetilde{\bQ}_I$ at $-\ii \infty$. This means that $\mathcal{L}_I{}^J$ is upper-triangular. We also require that the ordering is preserved at $u=-\infty$ which determines the exponential terms on the diagonal. The square-root cuts further add the restriction $\mathcal{L}_I{}^J\underline{\mathcal{L}}_J{}^K=\delta_I^K$ constraining the coefficients on the diagonal. Finally, we require that we do not add terms with different powers of $u$ in their asymptotics. Supplementing these restrictions with the weak coupling algorithm in the next section we found the matrix relation
\begin{equation}\label{eq:gluingG}
    \widetilde{\bQ}_I = {\begin{pmatrix}
        e^{-2\pi u} & 0 & l_{13} & 0 & 0 \\
        0 & -e^{-2\pi u}  & 0 & l_{24} & 0 \\
        0 & 0 & -e^{2\pi u} &0 & 0 \\
        0 & 0 & 0 & e^{2\pi u} & 0 \\
        0 & 0 & 0 & 0 & \bullet
        \end{pmatrix}_{I}}^{\ J}\,\underline{\bQ}_J\,.
\end{equation}
The explicit form of $\mathcal{L}$ depends on $H$-rotations. For the choices described in this section we find that $l_{13} = -l_{24} = 16\ii \sinh^2{\frac{1}{2 T_H}}$.

\subsection{Perturbative solution}\label{sec:PertQSCSol}

We now turn to the question of solving the constraints described in the previous section to obtain $T_H$.  In this section we consider the case at weak coupling, while in section~\ref{sec:StrongCoupling} we consider strong coupling.

The first step in solving the ABJM QSC at weak coupling is to find the Q-system at zero coupling. At $\coup=0$ we expect that the Q-system should be fully regular, that is, $Q_{a|i}$ should be polynomials in $u$ with exponential prefactors. To emphasize that we are now working at $\coup=0$ we write $y_0^{-\ii u}=e^{-\pi u} \,\partitionY_H^{\ii u}, \ \partitionY_H =\exp(-\frac{1}{2T_H^{(0)}})$. The polynomial structure of the Q-functions at $\coup=0$ can be anticipated both from the knowledge of how to treat a twisted spin-chain using Q-functions \cite{Kazakov:2015efa} and from the Hagedorn QSC for AdS$_5 \times$S$^5$ \cite{Harmark:2021qma}. To fix the exact form of the Q-functions we use that the ABJM QSC requires that \cite{Bombardelli:2017vhk}
\begin{align*}
     Q_{A|IJ}&=(\overline{\sigma}_A)^{ab} Q_{a|i}^+\kappa^{ij}(\Sigma_{IJ})_j{}^k Q_{b|k}^-=-(\sigma_A)_{ab}(Q^{b}{}_{|i}{})^-\kappa^{ij}(\Sigma_{IJ})_j{}^k (Q^{a}{}_{|k}{})^+\,,\\
     Q_{AB|I}&=(\sigma_{AB})_a{}^b(Q^a{}_{|i}{})^+ (\overline{\Sigma}_I)^{ij} Q_{b|j}^-=(\sigma_{AB})_a{}^b(Q^a{}_{|i}{})^- (\overline{\Sigma}_I)^{ij} Q_{b|j}^+\,.
\end{align*}
as well as \eqref{eq:InverseEq}. In addition, we use the last $H$-symmetry to impose the condition $\bQ_{1}\propto u$. This fixes all  constants in the {\it ansatz}, leaving  only the Hagedorn temperature to be fixed. We collect the explicit expressions for $\bP$ and $\bQ$ in Appendix~\ref{app:TreeLevel}. 

To fix $T_H$ we compute $\widetilde{\bQ}_I$ using \eqref{eq:gluingG}. Since $\widetilde{\bQ}_I+\bQ_{I}$ and $\frac{\widetilde{\bQ}_{I} - \bQ_{I}}{\sqrt{u-2\coup}\sqrt{u+2\coup}}$ do not have cuts on the real axis, they must be regular functions around $u\simeq 0$, that is
\begin{align}\label{eq:regConditions}
    &\widetilde{\bQ}_{I} + \bQ_{I}\bigg|_{u\simeq 0} = \text{reg}\,, 
    &
    &\frac{\widetilde{\bQ}_{I} - \bQ_{I}}{\sqrt{u-2\coup}\sqrt{u+2\coup}}\bigg|_{u\simeq 0} = \text{reg}\,. 
\end{align}
Imposing these relations at $\coup=0$ implies that $T_H^{(0)} =\frac{1}{2\log(3+2\sqrt{2})}$, which we recognize as a solution to $Z_{bead} = 1$ \footnote{The other solution to $Z_{bead}=1$ results in a negative temperature which we discard.}. This is how the QSC finds the Hagedorn temperature at $h=0$.

\subsubsection{Going to higher orders}\label{higher}
Techniques to solve the AdS$_4$ QSC perturbatively at small coupling were developed in \cite{Anselmetti:2015mda,Bombardelli:2018bqz}, see also \cite{Lee:2017mhh,Lee:2018jvn,Lee:2019oml}. We will introduce a slightly different algorithm inspired by \cite{Gromov:2015vua}.

To describe our perturbative algorithm we start by rewriting \eqref{eq:PQfromQai} as
\begin{equation}\label{eq:PerturbativeQai}
    Q_{a|i}^+ + \bP_{ab}\kappa^{bc}Q^-_{c|j} \ds{K}^j_i = 0\,,
\end{equation}
where $\bP_{ab} = \bP_A\, \sigma^{A}_{ab}$. Let $Q^{(m)}_{a|i}$ solve \eqref{eq:PerturbativeQai} to order $\mathcal{O}(h^{2m})$ and write
\be\label{eq:Qpert}
    Q_{a|i} = Q_{a|i}^{(m)} + h^{2m+2}Q_{a|j}^{(0)}(b^j_i)^++\mathcal{O}(h^{2m+4})\,.
\ee
Using \eqref{eq:InverseEq} we can solve for $b^{i}_j$ as
\begin{equation}\label{eq:FiniteDiffB}
    b^i_j - (-1)^{\frac{i(i-1)+j(j-1)}{2}}(b^{[2]})^i_j=(dS^{(m+1)})^i_j\,,  
\end{equation}
where
\begin{align}
   &(dS^{(m+1)})^i_j = -\ds{K}_{k}^{i}\kappa^{kl}((Q^{a}{}_{|l})^{(0)})^+\bP_{ab}\kappa^{bc}(Q_{c|j}^{(m)})^-\bigg|_{h^{2m+2}}\,.
\end{align}
We then notice that $(dS^{(m)})^{k}_{l} = y_0^{-\ii u\floor{\frac{k}{2}}+\ii u \floor{\frac{l}{2}}}(ds^{(m)})^{k}_l$\,, where $ds^{(m)}$ does not contain any exponential prefactors. If we now choose the  parameterization $b^{k}_l = y_0^{-\ii u\floor{\frac{k}{2}}+\ii u \floor{\frac{l}{2}}}p^{k}_l$\,, then \eqref{eq:FiniteDiffB} becomes
\begin{align}\label{eq:finiteDiffP}
    &p^k_l - z \left(p^{k}_l\right)^{[2]} = (ds^{(m+1)})^k_l  \,,
    &
    &z \in \{1,\partitionY_H,\partitionY_H^2,\frac{1}{\partitionY_H},\frac{1}{\partitionY_H^2}\}\,,
\end{align}
where  $\partitionY_H$ is given in \eqref{wH}. It is interesting to note that the additional signs appearing in \eqref{eq:FiniteDiffB} cancel against signs coming from $e^{\pm \pi u}$. 

The technology for solving equations of the type appearing in \eqref{eq:finiteDiffP} was developed in \cite{Leurent:2013mr,Marboe:2014gma,Gromov:2015dfa}. Let us briefly recall the procedure.
In the first iteration the source-term is $ds^{(1)}$, which from our parameterization  is a Laurent polynomial in $u$. For positive powers of $u$ it is straightforward to find $p$ as a polynomial in $u$. For negative powers we need the twisted $\eta$-function $\eta^{z}_{s}$ \cite{Gromov:2015vua}, which are functions without poles in the upper half-plane that solve
\begin{equation}
    \eta^{z}_s - z \,(\eta^{z}_s)^{[2]} = \frac{1}{u^s}\,.
\end{equation} 
A formal expression is given by
\begin{equation}
    \eta^{z}_s = \sum_{n=0}^{\infty} \frac{z^n}{(u+\ii n)^{s}}\,.
\end{equation}
Going to $\mathcal{O}(h^{4})$  will require $\eta^{z}_s$ appearing on the right-hand side of \eqref{eq:finiteDiffP}. To solve equations of this type we need generalized twisted $\eta$-functions, which are defined to satisfy
\begin{equation}\label{eq:FiniteDifferenceEta}
    \eta_{s,S}^{z,Z}-zZ\left(\eta_{s,S}^{z,Z}\right)^{[2]} = \frac{Z}{u^{s}} (\eta^{Z}_{S})^{[2]}\,,
\end{equation}
without poles in the upper half-plane and with $S$ and $Z$ multi-indices.
A formal expression for $\eta^{z_1,z_2,\dots,z_k}_{s_1,s_2,\dots,s_k}$ is
\begin{equation}\label{eq:TwistedEta}
    \eta_{s_1,s_2,\dots,s_k}^{z_1\, z_2,\dots z_k} = \sum_{0 \leq n_1 < n_2 < \dots < n_k<\infty} \frac{z_1^{n_1}}{(u+\ii n)^{s_1}}\frac{z_2^{n_2}}{(u+\ii n_2)^{s_2}}\dots \frac{z_k^{n_k}}{(u+\ii n_k)^{s_k}}\,.
\end{equation}
There is no need to introduce any more functions.

We note that there are also homogeneous solutions to \eqref{eq:finiteDiffP}. For example, when $z=1$ we can always add an $\ii$-periodic function to $p$ and still solve $\eqref{eq:finiteDiffP}$. However, since by construction $p$ must have polynomial asymptotic behavior and cannot have poles in the upper-half plane, we can discard all homogeneous solutions except for constants. We thus write
\begin{equation}
    p_{l}{}^{k} = \Psi_z\left((ds^{(m)})^{k}_l\right) + \delta_{\floor{\frac{k}{2}},\floor{\frac{l}{2}}} (p_H)^{k}_l\,,
\end{equation}
with $(p_H)^k_l$ constants and $\Psi_z$ a particular solution of \eqref{eq:finiteDiffP} analytic in the upper half-plane, {\it i.e.} written in terms of $\eta$-functions.

Thus, at each order in $\coup$ we can find $Q_{a|i}$ as a function of $\eta$-functions and $u$ with undetermined coefficients $c_{A,m}$ and $p_{H}$. To fix these coefficients we follow the strategy in \cite{Harmark:2021qma}. The first step is to construct $Q^{a}{}_{|i}$ using \eqref{eq:SymmetriSectorQcs} and then impose \eqref{eq:InverseEq}. The second step is to verify the asymptotics \eqref{eq:AsymptoticQP} for $\bQ_I$. In our weak-coupling expansion 
\begin{equation}
    T_H = \sum_{m=0}^{\infty} T_H^{(m)} \, h^{2m}\,, 
\end{equation}
so that
\begin{equation}
    y^{\ii u} = e^{\pi u}e^{\frac{\ii u}{2T_H^{(0)}}}\left(1-\ii u\,\frac{T_H^{(1)}}{2(T_H^{(0)})^2}\, \coup^2\right) + \mathcal{O}(\coup^4)\,.
\end{equation}
It follows that $\bQ_I$, and all other functions containing $y^{-\ii u}$, when expanded first in $\coup$, will look like they contain higher powers of $u$ compared to \eqref{eq:AsymptoticQP}. To make sure we still have the correct asymptotic behavior we solve \cite{Harmark:2021qma} \begin{equation}
    \frac{\bQ_{1}}{\bQ_{2}} \simeq u\,.
\end{equation}

The final step in the algorithm is to impose \eqref{eq:regConditions}. When doing so it is necessary to expand the $\eta$-functions close to $u=0$. As can be seen from \eqref{eq:TwistedEta}, this will introduce the polylogarithms
\begin{align}
    \text{Li}_{s_1,\dots,s_k}(z_1,z_2,\dots,z_k) = \sum_{1\leq n_1<n_2<\dots <n_k} \frac{z_1^{n_1}\dots z_k^{n_k}}{n_1^{s_1}\dots n_k^{s_k}}\,,
\end{align}
with $z_i \in \{1,\partitionY_H,\partitionY_H^2,\partitionY^{-1}_H,\partitionY^{-2}_H\}$. After ensuring that \eqref{eq:regConditions} is satisfied we can read off $T_H$ from \eqref{eq:tAtBConditions}.

\subsubsection{The Hagedorn temperature to order $\mathcal{O}(h^8)$}
Using the algorithm described in section~\ref{higher} we computed the Hagedorn temperature up to $\mathcal{O}(h^8)$. %Explicit exact expressions can be found in an ancillary file attached to the arXiv submission.
The results are (using $\partitionY_H=3-2\sqrt{2}$)
\begin{align}
    T_H^{(0)} =& \frac{1}{2 \log \left(3+2 \sqrt{2}\right)}\approx 0.283648164276628\,,
\end{align}
\begin{align}
    T_H^{(1)} =& \frac{\sqrt{2}-1}{\log \left(1+\sqrt{2}\right)}\approx 0.469963666342443\,,
\end{align}
\begin{equation}
    \begin{split}
        T_H^{(2)}=&7 \sqrt{2}-8 -4 \left(1+2 \sqrt{2}\right) \text{Li}_{1}\left(\frac{1}{\left(1+\sqrt{2}\right)^2}\right)-\frac{2 \left(1+2 \sqrt{2}\right)\text{Li}_{2}\left(\frac{1}{\left(1+\sqrt{2}\right)^2}\right)}{\log \left(1+\sqrt{2}\right)}  \\
        \approx&-2.54281120753405\,,
        \end{split}
\end{equation}
\begin{align}
    \begin{split}
        T_H^{(3)}=&\frac{4}{3} \left(48 \text{Li}_{1,1}\left(\partitionY_H,\partitionY_H\right)-48 \text{Li}_{1,1}\left(\partitionY_H^{-1},\partitionY_H\right)\right.\\
        &+12 \sqrt{2} \left(\text{Li}_{2}\left(\partitionY_H^2\right)-2 \log \left(\partitionY_H\right)\text{Li}_1\left(\partitionY_H^2\right)\right)\\
        &+(72 \sqrt{2}+48)\left(- \frac{\log \left(\partitionY_H\right)}{6}\text{Li}_{2}\left(\partitionY_H\right)+\frac{1}{2}\text{Li}_{3}\left(\partitionY_H\right)-\frac{\text{Li}_{4}\left(\partitionY_H\right)}{2\log\left(\partitionY_H\right)} \right)\\
        &+\text{Li}_{1}\left(\partitionY_H\right) \left(48 \sqrt{2}\text{Li}_{2}\left(\partitionY_H\right)-9 \left(4-5 \sqrt{2}\right) \log \left(\partitionY_H\right)\right)\\
        &-24 \sqrt{2}\log\left(\partitionY_H\right)\left(\text{Li}_{1}\left(\partitionY_H\right)\right)^2-\frac{6\sqrt{2}\left(\text{Li}_{2}\left(\partitionY_H\right)\right)^2}{\log \left(\partitionY_H\right)}\\
        &\left.+(-45 \sqrt{2}+84)\text{Li}_{2}\left(\partitionY_H\right)-\frac{1}{2}(35 \sqrt{2}-52)\log \left(\partitionY_H\right)+45 \sqrt{2}-66\right)
    \end{split}\\
    \approx& 21.7782105898884\nonumber\,,\\
    T_H^{(4)}\approx&-222.299692062791\,.
    \label{eq:PertSol}
\end{align}
The analytic expression for $T^{(4)}_H$ is given in appendix \ref{app:Th4}.
All numerical estimates have been produced using DiffExp \cite{Hidding:2020ytt}. We see that  our result for $T_H^{(0)}$ and $T_{H}^{(1)}$ agrees with the tree-level result of \cite{Nishioka:2008gz} and with the 2-loop calculation in \cite{Papathanasiou:2009en}, up to the aforementioned factor of $2$.  The results for $T_H^{(2)},T_H^{(3)}$ and $T_{H}^{(4)}$ are new.

When the spectral problem for AdS$_4$ was explored at finite coupling~\cite{Anselmetti:2015mda,Bombardelli:2018bqz}, a Padé approximation computed from the perturbative data was used to get a better agreement with the numerical results. For the Hagedron temperature we can do the same and find that the $[4/4]$ Padé approximation is \begin{equation}\label{eq:Pade}
    T_H^{\textrm{Padé}[4/4]}=\frac{0.283648164276628+4.16468030174843 h^2+10.4253195236392 h^4}{1+13.0257026158742 h^2+24.1373808661550 h^4}.
\end{equation}

\subsection{The numerical solution}\label{sec:StrongCoupling}
To go to strong coupling we will resort to numerics. We will use standard methods first introduced for $\mathcal{N}=4$ in \cite{Gromov:2015wca} and implemented for the spectral problem of ABJM in \cite{Bombardelli:2018bqz} (see \cite{Levkovich-Maslyuk:2011ccm} for earlier numerical results from TBA). However, due to our exponential asymptotics we are forced to work slightly harder compared to the spectral problem, just as in \cite{Harmark:2021qma}.

\subsubsection{The algorithm}
We can also use the {\it ansatz} in \eqref{eq:AnsatzP}  for the numerical solution. We only need to truncate all sums at  $n=K$ so as to work with a finite number of unknowns.  We then assume that for large $u$ that $Q_{a,i}$ takes the form
\begin{equation}
    Q_{a|i}=y^{s_i\ii u}u^{p_{a|i}}\sum_{n=0}^N b_{a|i}^{(n)}u^{-n}\,,
\end{equation}
where 
\begin{align*}
    &s_i=\begin{cases} -1,\ i=1\\ 1,\ i=4\\ 0,\ i=2,3\end{cases}
    &
    &p_{a|i}=\delta_{i,2}+\begin{cases}
        0 & a = 1\\
        2 & a = 2 \\
        1 & a = 3 \\
        3 & a = 4
    \end{cases}\,,
\end{align*}
and $N$ is an integer cut-off that satisfies $N=2K-2$.
The large $u$ expansion of $Q_{a|i}$ is constrained by the parity condition \eqref{eq:ParityQ}. Parity implies that $b_{a|2}^{(n)},b_{a|3}^{(n)}$ are zero for $n$ odd and
\begin{equation}
    b_{a|4}^{(n)}= \frac{\ii (-1)^n }{8 \sinh^3\frac{1}{2T_H}}b_{a|1}^{(n)}
\end{equation} 
for all $n$. Note that the asymptotic form of $\Omega$ in \eqref{eq:AsymptoticOmega} is exact only up to exponentially suppressed terms at large u, so we can impose these parity constraints only on the large-$u$ expansion. As a gauge choice we also set 
\begin{align}
    b_{1|1}^{(0)}&=2\cosh\left(\frac{1}{4T_H}\right), & b_{1|2}^{(0)}&=\frac{1}{2\sinh\left(\frac{1}{4T_H}\right)}\,.
\end{align}

With these simplifications we can start solving order by order in $u^{-1}$ the equation \begin{align}
    Q_{a|i}^++\bP_{ab}\kappa^{bc}Q_{c|j}^-\mathds{K}^j_i=0\,.
\end{align} At the leading five orders we solve for the following coefficients: 
\begin{align*}
    u^6&: & &b_{4|2}^{(0)},\\
    u^5&: & &b_{4|3}^{(0)},b_{3|2}^{(0)},b_{2|2}^{(0)},b_{4|1}^{(0)},\\
    u^4&: & &b_{4|2}^{(2)}, b_{3|3}^{(0)}, b_{2|3}^{(0)}, b_{4|1}^{(1)}, b_{2|1}^{(0)}, b_{3|1}^{(0)},\\
    u^3&: & &b_{4|3}^{(2)}, b_{3|2}^{(2)}, b_{2|2}^{(2)}, b_{4|1}^{(2)}, b_{2|1}^{(1)}, b_{3|1}^{(1)},\\
    u^2&: & &b_{4|2}^{(4)}, b_{3|3}^{(2)}, b_{2|3}^{(2)}, b_{4|1}^{(3)}, b_{2|1}^{(2)}, b_{3|1}^{(2)}, c_{5,0}, c_{4,-1}.
\end{align*}

At this point it is convenient to solve a constraint coming from the restriction to the symmetric sector. Combining \eqref{eq:SymmetriSectorQcs} and \eqref{eq:InverseEq} leads to
\begin{align}
    \kappa^{ij}Q_{a|j}(-\kappa^{bc}Q_{c|k}\mathds{K}^{k}_i)=\delta_a^b\,.
\end{align}
Expanding this to  order $u^4$ then fixes $b_{1|3}^{(0)}$. Now we can go back to the main equation where we solve the next three orders: 
\begin{align*}
    u^1&: & &b_{4|3}^{(4)}, b_{3|2}^{(4)}, b_{2|2}^{(4)}, b_{4|1}^{(4)}, b_{2|1}^{(3)}, b_{3|1}^{(3)},\\
    u^0&: & &b_{4|2}^{(6)}, b_{3|3}^{(4)}, b_{2|3}^{(4)}, b_{4|1}^{(5)}, b_{2|1}^{(4)}, b_{3|1}^{(4)}, c_{4,1}, c_{3,0},\\
    u^{-1}&: & &b_{4|3}^{(6)}, b_{3|2}^{(6)}, b_{2|2}^{(6)}, b_{4|1}^{(6)}, b_{2|1}^{(5)}, b_{3|1}^{(5)}, b_{1|1}^{(1)}.
\end{align*}
After this, for all $k\geq 1$ we can solve in order  for: \begin{align*}
    u^{-2 k}&: & &b_{4|2}^{(6+2k)}, b_{3|3}^{(4+2k)}, b_{2|3}^{(4+2k)}, b_{4|1}^{(5+2k)}, b_{2|1}^{(4+2k)}, b_{3|1}^{(4+2k)}, b_{1|1}^{(2k)}, b_{1|2}^{(2k)}, c_{3,2k},\\
    u^{-(2k+1)}&: & &b_{4|3}^{(6+2k)}, b_{3|2}^{(6+2k)}, b_{2|2}^{(6+2k)}, b_{4|1}^{(6+2k)}, b_{2|1}^{(5+2k)}, b_{3|1}^{(5+2k)}, b_{1|1}^{(1+2k)}, b_{1|3}^{(2k)}.
\end{align*}

The aim now is to check that $\bQ_I$ glues correctly on the cut along the real line at the points 
$$I={\left\{-2h\cos\left(\frac{\pi}{I_P}\left(n-\frac{1}{2}\right)\right)\right\}_{ n=1,\dots,I_P}}\,.$$ 
To get there we can start for each $v\in I$ with $\bQ_{a|i}$ from $u=v+\ii \frac{U}{2}$ for $U$ a large odd positive integer and iterate using the equation \begin{align}
    Q_{a|i}^-&=-\bP_{ab}\kappa^{bc}Q_{c|j}^+\mathds{K}^j_i
\end{align} to get $\bQ_{a|i}^+(v)$. Then we can compute $\bQ_I$ and $\widetilde{\bQ}_I$ on opposite sides of the cut from \begin{align}
    \bQ_I(v+\ii 0)&=-\frac{1}{2}\left(\overline{\Sigma}_I\right)^{ij}\kappa^{ab}Q_{b|k}^+(v)\mathds{K}^k_i\bP_{ac}(v+\ii 0)\kappa^{cd}Q_{d|l}^+(v)\mathds{K}^l_j\\
    \widetilde{\bQ}_I(v-\ii 0)&=\mathcal{L}_I{}^J(x)\left(-\frac{1}{2}\left(\overline{\Sigma}_J\right)^{ij}\kappa^{ab}Q_{b|k}^+(-v)\mathds{K}^k_i\bP_{ac}(-v+\ii 0)\kappa^{cd}Q_{d|l}^+(-v)\mathds{K}^l_j,\right)
\end{align}
with $\mathcal{L}$ the gluing matrix from \eqref{eq:gluingG}. We can also get $\widetilde{\bP}$ from $\bP$ by sending the Zhukovsky variable $x$ to $\frac{1}{x}$ and thus get tilded versions of the above expressions by replacing $\bP$ with $\widetilde{\bP}$. Continuity through the cut implies that for the exact solution the function \begin{equation}
    F(\{ c_{A,n}\},y)=\sum_{v\in I}\sum_{J=3}^4\left(\left\vert \frac{\bQ_J(v+\ii 0)}{\widetilde{\bQ}_J(v-\ii 0)}-1 \right\vert^2+\left\vert \frac{\widetilde{\bQ}_J(v+\ii 0)}{{\bQ}_J(v-\ii 0)}-1 \right\vert^2 \right)
\end{equation} is zero. We can then use  the Levenberg-Marquardt algorithm \cite{Gromov:2015wca} to find an approximate numerical solution.

\subsubsection{The results}\label{results}
For weak coupling in the range  $0<h^2\le0.1$ we use the values $K=15$, $N=28$, $U=91$ and $I_P=64$. For  strong coupling with $h\le1.99$ we use the same values. For $1.99<h\le 3.99$ we use $K=20$, $N=38$, $U=91$ and $I_P=84$, for $4\leq h\leq 4.44$ we use $K=21$, $N=40$, $U=121$ and $I_P=88$, while for $4.48\leq h$ we increase $K$ and $N$ to $K=25$, $N=48$. We provide all  results for the Hagedorn temperature at the above range of couplings in the ancillary files {\it weak\_h\_squared.csv} and {\it strong\_h.csv}.

To check the numerical accuracy, we perturbed the result at $h^2=0.1$ and reran the algorithm 21 times at this same coupling to find that the Hagedorn temperature  converged to the same value with an accuracy of 41 digits on  all 21 runs. At the strong coupling point $h=3.99$ we did the same for 10 runs with perturbations of order $10^{-6}$ and two runs at order $10^{-5}$ to find results agreeing with 12 digit accuracy.

\begin{figure}
    \centering
    \includegraphics[width=\textwidth]{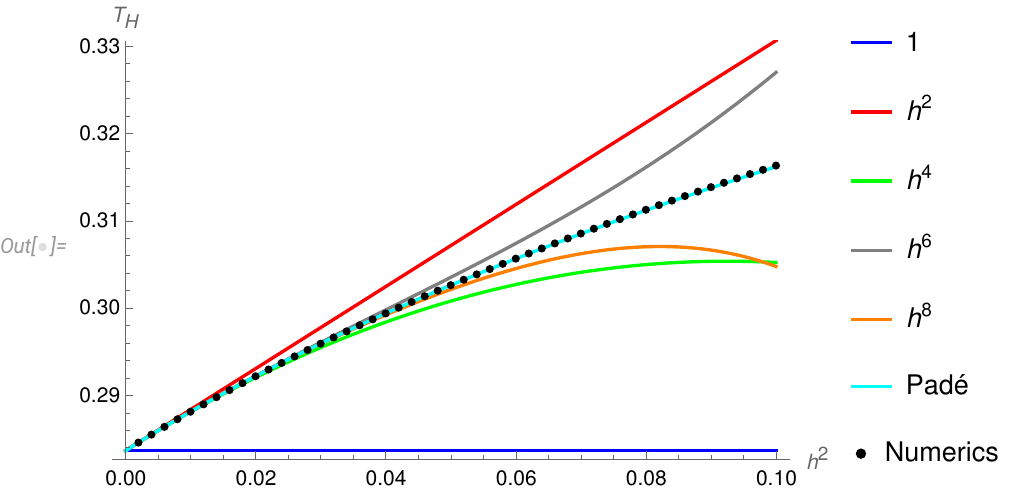}
    \caption{The perturbative solution of the QSC compared to the numerical solution at weak coupling. In the legend we indicate the order in $h$ of the perturbative expansion. The Padé approximation \eqref{eq:Pade} fits the numerical data to a relative difference of $4\times 10^{-4}$ across the full range shown in the plot.}
    \label{fig:WeakCoupling}
\end{figure}

At weak $h$ the numerics agrees well with the perturbative solution of the QSC, as shown in figure \ref{fig:WeakCoupling}. By fitting the numerical data with an even polynomial of order $h^{98}$ we match the first five terms with 15 digit precision against their exact values \eqref{eq:PertSol}. In the plot we can observe the break down of the perturbative expansion in $h^2$ before $h^2=0.1$. However the Padé approximation \ref{eq:Pade} fits well on the whole range of the weak coupling data.

\begin{figure}
    \centering
    \includegraphics[width=\textwidth]{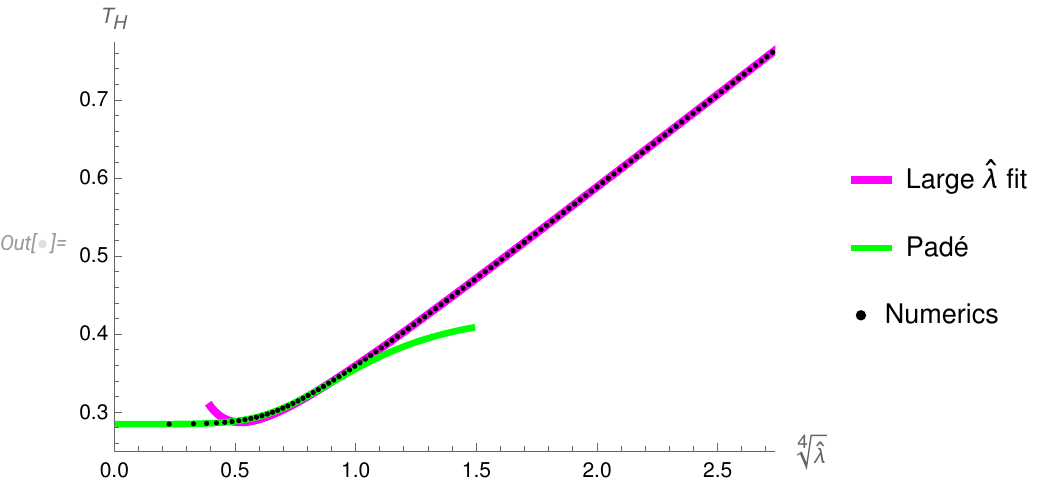}
    \caption{We show (half of) the numeric results for the Hagedorn temperature up to $\hatlambda^{1/4}\approx2.74$. The strong coupling fit matches within a relative error of less than one tenth of a percent all the way down to $\hatlambda^{1/4}\approx 0.83$, while the Padé approximation \eqref{eq:Pade} fits within the same error margins all the away up to $\hatlambda^{1/4}\approx 0.88$.}
    \label{fig:NumericsWeakTo Strong}
\end{figure}

At strong coupling we convert from $h$ to $\hatlambda=\lambda-\frac{1}{24}$ using \eqref{largelam} and fit against a series in negative powers in $\hatlambda^{1/4}$. To mitigate the errors  choosing the correct range of data to fit on, we average over different values for the lowest value of $\hatlambda$ and choose the range which gives the smallest  estimated standard deviation for the leading fitted coefficient. We also vary how many orders we include in the series. This way we can check that the large $\hatlambda$ asymptotics agrees with the string theory prediction at orders $\hatlambda^{1/4}$ and $\hatlambda^0$. To get better numerical estimates for the subleading terms, we then subtract the exact string theory prediction for the two leading terms from our numerical data. Taking the mean of the resulting coefficients and estimating the standard deviation we find \begin{multline}
    T_H(\hatlambda)=\frac{\hatlambda^{1/4}}{2^{5/4}\sqrt{\pi}}+\frac{3}{8\pi}\\
    -(0.0308\pm 0.0004)\hatlambda^{-1/4}+(0.046\pm 0.003)\hatlambda^{-1/2}-(0.017\pm 0.006)\hatlambda^{-3/4}\\
    +(0.005\pm0.004)\hatlambda^{-1}+(0.0004\pm0.0006)\hatlambda^{-5/4}+\mathcal{O}(\hatlambda^{-3/2}),
\end{multline} 
We plot the numerical data and the fitted function in figure \ref{fig:NumericsWeakTo Strong}. We observe that  the $\hatlambda^{-1/2}$ term matches  the prediction in \eqref{THconj} within the error margins.  If we now assume that the predicted value for this coefficient is correct,  we can  make a more precise estimate for the $\hatlambda^{-1/4}$ coefficient, and hence the value for $\Delta c$,  by refitting the curve.  Our result is 
\begin{equation}\label{eq:conjecture}
    \Delta c=-2.0782\pm 0.0016\approx -3\log(2)=2.07944\dots\,.
\end{equation}

In \cite{Ekhammar:2023glu} we  found that $\Delta c=-4\log(2)$ for AdS$_5$ with  slightly better numerical accuracy. This along with \eqref{eq:conjecture} leads us to conjecture that for general AdS$_{d+1}$ the contribution from the zero-point shift is \begin{equation}\label{deltcconj}
    \Delta c=-d\log(2).
\end{equation}

%%%%%%%%%%%%%%%%%%%%%%%%%%%%%%%%%%%%%%%
\section{Discussion}
%%%%%%%%%%%%%%%%%%%%%%%%%%%%%%%%%%%%%%%

In this paper we computed the Hagedorn temperature at weak and strong coupling for ABJM using the QSC.  At weak coupling we found the result up to  eight loops, although there is theoretically no limit as to how high we could go.  At strong coupling we found the Hagedorn temperature numerically in inverse powers of $\lambda^{1/4}$ to four nontrivial orders in the expansion, showing that the last two terms are consistent with a conjectured form for the coefficients in the expansion.

Putting equations \eqref{THd} and \eqref{eq:conjecture} together we  observe  a linear dependence  on the dimension of the CFT for $\Delta c$. However one could twist the $R$-symmetries by turning on fugacities and explore how deformations related to the internal space affect the asymptotic behavior. In this case there is less symmetry in the QSC so the numerical calculations are slower.  Perhaps recent work on numerical solutions of the QSC would be of use \cite{Gromov:2023hzc}. This investigation is in progress.

It would also be interesting to study the AdS$_3$ case. The first correction in the expansion was already given in \cite{Urbach:2023npi} and verified on the world-sheet in \cite{Bigazzi:2023oqm},  with the next two terms given in \eqref{THconj} and \eqref{deltcconj}. For AdS$_3$ there is a conjectured form for the QSC \cite{Ekhammar:2021pys,Cavaglia:2021eqr} and a TBA \cite{Frolov:2021bwp}.  However, even for the spectral problem, it is not known how to take the AdS$_3$ QSC to strong coupling with the present technology due to the novel analytic properties of the curve.

%%%%%%%%%%%%%%%%%%%%%%%%%%%%%%%%%%%%%%%
\section*{Acknowledgements}
%%%%%%%%%%%%%%%%%%%%%%%%%%%%%%%%%%%%%%%%

We thank 
Martijn Hidding for help with the numerical evaluation of polylogarithms.  We also thank N. Bobev, N. Gromov, R. Tateo, G. Papathanasiou and M. Spradlin for discussions.
This research is supported in part by the Swedish Research Council under grant \#2020-03339 and by the National Science Foundation under Grant No.~NSF PHY-1748958. Computations were done on a  cluster provided by the National Academic Infrastructure for Supercomputing in Sweden (NAISS) at UPPMAX, partially funded by the Swedish Research Council under grant \#2022-06725. S.E was partially supported by the European Research Council
(ERC) under the European Union’s Horizon 2020 research and innovation programme (grant
agreement No. 865075) EXACTC.
J.A.M. thanks %the Center for Theoretical Physics at MIT and 
the KITP for hospitality during the beginning of this work.

\appendix
\section{ABJM Conventions}\label{app:Conventions}
We follow the conventions of \cite{Bombardelli:2017vhk}. We use $A,B=1,\dots,6$ for $SO(6)$ vector indices, and $a,b=1,\dots,4$ for the corresponding spinor indices. Relevant matrices pertaining to the $SO(6)$ are
\begin{align}
    &\eta_{AB} = \begin{pmatrix}
        0 & 0 & 0 & 1 & 0 & 0 \\
        0 & 0 & -1 & 0 & 0 & 0\\
        0 & -1 & 0 & 0 & 0 & 0 \\
        1 & 0 & 0 & 0 &0 & 0 \\
        0 & 0 & 0 & 0 & 0 & 1 \\
        0 & 0 & 0 & 0 & 1 & 0
    \end{pmatrix}\,,
    &
    &\kappa_{ab}= \begin{pmatrix}
        0 & 0 & 0 & 1 \\
        0 & 0 & -1 & 0 \\
        0 & 1 & 0 & 0 \\
        -1 & 0 & 0 & 0 
    \end{pmatrix}\,,\\
    &V_A \sigma^A_{ab} = \begin{pmatrix}
        0 & -V_1 & -V_2 & -V_5 \\
        -V_1 & 0 & -V_6 & -V_3 \\
        V_2 & V_6 & 0 & -V_4 \\
        V_5 & V_3 & V_4 & 0
    \end{pmatrix}\,, 
    &
    &V_A (\bsigma^A)^{ab} = \begin{pmatrix}
        0 & V_4 & -V_3 & V_6 \\
        -V_4 & 0 & V_5 & -V_2 \\
        V_3 & -V_5 & 0 & V_1 \\
        V_5 & V_3 & -V_1 & 0
    \end{pmatrix}\,.
\end{align} The $SO(5)$ conventions are derived from the above $SO(6)$ conventions by identifying components 5 and 6. We use $I,J=1,\dots5$ for the vector and $i,j=1,\dots 4$ for the spinor indices of the $SO(5)$.  Relevant matrices are 
\begin{align}
    & \rho_{IJ}=\begin{pmatrix}
        0 & 0 & 0 & 1 & 0\\
        0 & 0 & -1 & 0 & 0\\
        0 & -1 & 0 & 0 & 0\\
        1 & 0 & 0 & 0 & 0\\
        0 & 0 & 0 & 0 & \frac{1}{2}
    \end{pmatrix}\,,
\end{align}
\begin{align}
    &\Sigma^I=(\sigma^1,\sigma^2,\sigma^3,\sigma^4,\sigma^5+\sigma^6),
    &
     &\overline{\Sigma}^I=(\overline{\sigma}^1,\overline{\sigma}^2,\overline{\sigma}^3,\overline{\sigma}^4,\overline{\sigma}^5+\overline{\sigma}^6)\,,\\
    &\kappa_{ij}= \begin{pmatrix}
        0 & 0 & 0 & 1 \\
        0 & 0 & -1 & 0 \\
        0 & 1 & 0 & 0 \\
        -1 & 0 & 0 & 0 
    \end{pmatrix}\,,
    &
    & \mathds{K}^i_j = \begin{pmatrix}
        1 & 0 & 0 & 0\\
        0 & -1 & 0 & 0\\
        0 & 0 & -1 & 0\\
        0 & 0 & 0 & 1
    \end{pmatrix}\,.
\end{align}

\section{Tree level solution}\label{app:TreeLevel}

For the tree-level solution, $\coup=0$, we find the following expressions for $\bP$ and $\bQ$
\begin{align}
    \bP_A&=\begin{pmatrix}
        u\\
        1\\
        -\frac{4}{3}u^4\cosh^4 \frac{1}{4T} +\frac{\left(\cosh^2\frac{1}{4T}-6\right)^2}{9}-1\\
        -\frac{16}{3}u^3\cosh^4\frac{1}{4T}-\frac{4}{3}\cosh^2 \frac{1}{4T}\left(\cosh^2\frac{1}{4T}-6\right)u\\
        2 u^2\cosh^2 \frac{1}{4T}+\frac{1}{3}\left(\cosh^2\frac{1}{4T}-6\right)\\
        2u^2\cosh^2\frac{1}{4T}+\frac{1}{3}\left(\cosh^2\frac{1}{4T}-6\right)
    \end{pmatrix}\,,\\
    \bQ_I&=\begin{pmatrix}
        8 \ty^{-\ii u} u \cosh^3\frac{1}{4T}\\
        8 \ty^{-\ii u}\cosh^3 \frac{1}{4T} \\
       \ty^{\ii u}\frac{1}{8\sinh^3\frac{1}{4T}}\left(\ii u+\frac{\sinh^2\frac{1}{4T}-1}{\sinh\left(\frac{1}{4T}\right)\cosh\left(\frac{1}{4T}\right)}\right)\\
        \ty^{\ii u}\frac{\ii}{8\sinh^3\frac{1}{4T}}\\
        0
    \end{pmatrix}\,,
\end{align}

\section{$T_H^{(4)}$}\label{app:Th4}
Using $w_H=3-2\sqrt{2}$ from section \ref{subsec:WeakCoupling} we find:
\begin{equation*}
    \begin{split}
    T_H^{(4)}=&56-296 w_H+4 \left(104 w_H-\frac{49}{3}\right) \log \left(w_H\right)+\frac{1}{4} \left(-331 w_H-31\right) \log ^2\left(w_H\right)\\
   +&64\left( \log ^2\left(w_H\right) \left(w_H-3\right)-3 \log \left(w_H\right)\right) \text{Li}_1\left(w_H\right){}^3\\
    +&8\left(3\left(13-7 w_H\right) \log ^2\left(w_H\right)+\left(9 w_H+5\right) \log \left(w_H\right)-4 \left(w_H-1\right)\right.\\
   &\left.+24\left(1-\log \left(w_H\right) \left(w_H-3\right)\right) \text{Li}_2\left(w_H\right)\right)\text{Li}_1\left(w_H\right){}^2\\
   +& 4\left(48 \left(w_H-3\right) \text{Li}_2\left(w_H\right){}^2+4 \left(w_H-11\right) \text{Li}_1\left(w_H^2\right) \log ^2\left(w_H\right)\right.\\
   &+\text{Li}_2\left(w_H\right) \left(75 w_H+4 \left(4 w_H-11\right) \log \left(w_H\right)-97\right) \log \left(w_H\right)\\
   &\left.-8 \left(w_H-1\right) \text{Li}_2\left(w_H^2\right)\log \left(w_H\right)\right)\text{Li}_1\left(w_H\right)-8 \left(23 w_H+27\right)\text{Li}_4\left(w_H\right)\\
   -&\frac{64 \left(w_H-3\right) \text{Li}_2\left(w_H\right){}^3}{\log \left(w_H\right)}+2 \left(-93 w_H+8\left(11-4 w_H\right) \log \left(w_H\right)+311\right)\text{Li}_2\left(w_H\right){}^2 \\
   +& 16\left(\left(3 w_H-5\right) \text{Li}_2\left(w_H^2\right)+\left(5-\frac{13 w_H}{3}\right) \log ^2\left(w_H\right)\right)\text{Li}_2\left(w_H\right)\\
   +& 4\left(5 \left(\frac{2 w_H}{3}-3\right) \log \left(w_H\right)+2 \left(27 w_H-17\right)\right)\log \left(w_H\right)\text{Li}_3\left(w_H\right)\\
   +&100 \left(9-2 w_H\right)\left(\frac{2}{5}\log \left(w_H\right) \text{Li}_4\left(w_H\right)-\text{Li}_5\left(w_H\right)+\frac{ \text{Li}_6\left(w_H\right)}{\log \left(w_H\right)}\right)\\
   +&\left(\text{Li}_1\left(w_H\right) \log \left(w_H\right)-\text{Li}_2\left(w_H\right)\right) \left(384 \text{Li}_{1,1}\left(w_H,w_H\right)+\frac{32 \left(w_H+1\right)\text{Li}_3\left(w_H^2\right)}{\log \left(w_H\right)}\right.\\
   &+48 \left(4 w_H-11\right) \left(\frac{\text{Li}_4\left(w_H\right)}{\log\left(w_H\right)}-\text{Li}_3\left(w_H\right)\right)-4 \left(4 \zeta (2) \left(w_H+5\right)+63 w_H+19\right)\\
   &\left.-\frac{32 \zeta (3) \left(w_H+1\right)}{\log\left(w_H\right)}+\frac{16}{3} \left(14 w_H+19\right) \log \left(w_H\right)\right)\\
   +& \left(16 \left(17-3 w_H\right) \text{Li}_2\left(w_H\right) +32 \left(7-5 w_H\right) \log\left(w_H\right)-64 \left(w_H-3\right) \right)\log \left(w_H\right)\text{Li}_1\left(w_H^2\right)\\
   +&2  \left(-32 w_H+\left(27 w_H-145\right) \log ^2\left(w_H\right)+4\left(11 w_H-1\right) \log \left(w_H\right)\right)\text{Li}_2\left(w_H^2\right)\\
   +&16 \left(5-3 w_H\right) \text{Li}_3\left(w_H^2\right) \log \left(w_H\right)+4 \left(19 w_H-121\right) \text{Li}_4\left(w_H^2\right)\\
   +&\left(64 \left(w_H-1\right)+4 \left(11 w_H-65\right) \log ^2\left(w_H\right)-16\left(9 w_H+5\right) \log \left(w_H\right)\right) \text{Li}_{1,1}\left(w_H,w_H\right)\\
   +&8 \left(w_H-3\right) \log \left(w_H\right)\text{Li}_{1,2}\left(\frac{1}{w_H},w_H\right)-128 \log \left(w_H\right) \text{Li}_{1,2}\left(\frac{1}{w_H},w_H^2\right)\\
   +&8 \left(3 w_H+7\right) \log \left(w_H\right)\text{Li}_{2,1}\left(\frac{1}{w_H},w_H\right)+32 \left(w_H-3\right) \log \left(w_H\right) \text{Li}_{2,1}\left(w_H,w_H\right)\\
   +&32 \left(w_H-11\right)\left( \text{Li}_{1,3}\left(w_H,w_H\right)- \text{Li}_{1,3}\left(\frac{1}{w_H},w_H\right)+\frac{\log \left(w_H\right)}{2}  \text{Li}_{2,1}\left(\frac{1}{w_H},w_H^2\right)\right)\\
   +&8\left(41-3 w_H\right)\left( \text{Li}_{2,2}\left(\frac{1}{w_H},w_H\right)-\text{Li}_{2,2}\left(w_H,w_H\right)\right)\\
   +&16 \left(w_H+5\right) \log ^2\left(w_H\right) \text{Li}_{1,1}\left(\frac{1}{w_H},w_H^2\right)+384 \left(\text{Li}_{3,1}\left(\frac{1}{w_H},w_H\right)- \text{Li}_{3,1}\left(w_H,w_H\right)\right).
   \end{split}
\end{equation*}

\bibliographystyle{JHEP}
\bibliography{ref}

\end{document}